\documentclass[reprint,amsmath,amssymb,aps,superscriptaddress]{revtex4-2}
\usepackage{graphicx}
\usepackage{dcolumn}
\usepackage{bm}
\usepackage{xcolor}
\usepackage{float}
\usepackage{mathrsfs}
\begin{document} 
\title{Explosive death in coupled oscillators with higher-order interactions}
\pacs{Valid PACS appear here}  
\author{Richita Ghosh}
 \affiliation{Department of Physics, Central University of Rajasthan, Rajasthan, Ajmer-305 817, India}
\author{Umesh Kumar Verma}%
\affiliation{%
Complex Systems Lab, Department of Physics, Indian Institute of Technology Indore, Khandwa Road, Simrol, Indore-453 552, India
}%

\author{Sarika Jalan}
\affiliation{%
Complex Systems Lab, Department of Physics, Indian Institute of Technology Indore, Khandwa Road, Simrol, Indore-453 552, India
}%
\author{Manish Dev Shrimali}
\email{shrimali@curaj.ac.in}
\affiliation{Department of Physics, Central University of Rajasthan, Rajasthan, Ajmer-305 817, India}
\begin{abstract}
We investigate  the dynamical evolution of globally connected Stuart-Landau oscillators coupled through conjugate or dis-similar variables on simplicial complexes. We report a first-order explosive phase transition from oscillatory state to death state, with $2$-simplex (triadic) interactions, as opposed to the second-order transition with only $1$-simplex (dyadic) interactions. Moreover, the system displays four distinct homogeneous steady states in the presence of triadic interactions, in contrast to the two homogeneous steady states observed with dyadic interactions. We calculate the backward transition point analytically, confirming the numerical results and providing the origin of the dynamical states in the transition region. 
The study will be useful in understanding complex systems, such as ecological and epidemiological, having higher-order interactions and coupling through conjugate variables.
\end{abstract}

\maketitle

\paragraph{Introduction}
The dynamics of interacting systems are usually modeled using the concept of complex networks \cite{pastor2015epidemic}. The units of each system are denoted as nodes and the interactions by links joining the nodes form networks. Most of the studies on networks deal with pairwise interactions between the nodes. However, such a simplified version of pairwise interactions is not justified while dealing with systems such as those formed by coauthors of research articles, multiple species competing for food in an ecosystem, or phenomena of diseases and rumors in society, etc.  
In nature, systems such as those governed by social \cite{benson2016higher,centola2018experimental}, biological \cite{sanchez2019high,ritz2014signaling}, neuroscience \cite{petri2014homological,yu2011higher} or ecological dynamics \cite{grilli2017higher,bairey2016high} are reigned by group interactions. The 
mathematical framework of pairwise links in networks is insufficient to represent such systems. Simplicial complexes are one of the simplest candidates for encoding higher-order interactions mathematically. A simplex of $N$ dimensions can be conceptualized as a set of $(n+1)$ nodes. The node set $\mathscr{D}$ may be written as $\mathscr{D}=[x_0,x_1,....x_n]$ where $[x_0,x_1,....x_n]$ is the set of the vertices (nodes) of the simplex.  A one-dimensional simplex represents a link between two nodes, while a two-dimensional simplex represents a triangle (denoting interaction between three nodes). Hence, simplicial complexes may be considered topological constructions constituting simplices of different dimensions \cite{boccaletti2023structure}.  

Further, dynamical processes evolving on complex networks exhibit collective emergent phenomena such as synchronization \cite{pecora1990synchronization}, self-organization \cite{lehn2002toward}, oscillation quenching \cite{koseska2013oscillation} and chimera states \cite{abrams2004chimera}. Among these, oscillation suppression in dynamical systems is a well-documented and interesting behavior arising due to interactions in complex systems. The onset of suppression of oscillations has been observed through various coupling schemes such as mean-field diffusive \cite{verma2018first}, through conjugate \cite{karnatak2009synchronization} and environmental variables \cite{arumugam2016environmental}, and with attractive-repulsive links \cite{dixit2020static} in networks of both limit-cycle and chaotic oscillators.  

Suppression of oscillations can occur through two routes; amplitude death (AD) \cite{saxena2012amplitude} and oscillation death (OD) \cite{koseska2010parameter}. For AD, the coupled oscillators stabilize to a common steady state which is also a stable global fixed point of the system whereas in OD, they can settle to different fixed points referred to as in-homogeneous (IHSS) steady state via symmetry-breaking bifurcation or  to a same fixed point referred as homogeneous (HSS) steady states. The suppression of oscillations is relevant at various places, for  instance, in medical science in controlling diseases like Parkinson's and Alzheimer's. Other examples include neuron models \cite{ermentrout1990oscillator}, lasers \cite{wei2007amplitude}, and climate systems \cite{gallego2001decadal}. 

Further, the transition from oscillations to the quenching of oscillations may be either continuous and smooth or discontinuous and explosive. 
There have been extensive studies on first-order transition to synchronization referred to as explosive synchronization \cite{gomez2011explosive}, explosive percolation \cite{bastas2014explosive}, and explosive death \cite{verma2017explosive} on networks with dyadic interactions. Later, explosive death (ED) \cite{zhao2017restoration,zhao2018explosive,han2016amplitude,sharma2019time,zhao2018enhancing,ponrasu2018conjugate,dutta2023oscillation} and semi-explosive death \cite{sun2021explosive} were  observed in a network of conjugate-coupled  limit-cycle oscillators. The phenomenon of ED brings with it a hysteresis or bi-stable region, i.e., a region in the parameter space where the oscillatory and the steady states coexist. Such bi-stability has been observed in a number of physical \cite{herrero2000experimental} and chemical systems \cite{bar1985stable}. With the advent of studies on higher-order interactions, collective dynamics, such as synchronization \cite{skardal2021higher,parastesh2022synchronization} and chimera   \cite{kundu2022higher} have been reported for coupled dynamics on simplicial complexes. 
In explosive transitions, a system suddenly experiences a transition to synchronization from an incoherence state and vice-versa \cite{skardal2021higher}. Further, multilayer networks of simplicial complexes have exhibited multiple routes to explosive synchronization \cite{millan2020explosive,jalan2022multiple}. Recently, a sudden transition to anti-phase synchronization on adaptive simplicial complexes has also been explored \cite{kachhvah2022first}.  

In this study, we extend the mathematical framework of triadic interactions \cite{gambuzza2021stability} to induct conjugate coupling, or coupling through dis-similar variables \cite{karnatak2007amplitude}. We investigate the effect of such a coupling scheme on limit-cycle oscillators and discover a first-order phase transition from oscillatory to steady state in the presence of triadic interactions instead of the continuous second-order phase transition that we observe with only dyadic interactions. 
We perform linear stability analysis for all the fixed points and derive the condition for the transition point from oscillation to steady state to further confirm our numerical results. 
\paragraph{Model:}
	
 Our model system with $1$-simplex (pairwise or dyadic) and $2$-simplex (triadic) interactions with conjugate or dis-similar variables has been mathematically formulated in the following way: 
\begin{eqnarray}
		\dot{\mathbf{X_i}}&=& \mathbf{F}(\mathbf{X_i})  + \frac{\epsilon_p}{N} \sum_{j=1}^{N} C_{ij}\mathbf{f_p}(\mathbf{X_i},\mathbf{X^{*}_{j}})\nonumber\\
		&&+\frac{\epsilon_h}{N^2} \sum_{j=1}^{N}\sum_{k=1}^{N} D_{ijk}\mathbf{f_h}(\mathbf{X_i},\mathbf{X^{*}_j},\mathbf{X_{k}})
		\label{eq:eqmode1}  
	\end{eqnarray}
where $\mathbf{X^{*}_{i}}$ is the conjugate variable of the state variable $\mathbf{X_i}$. The $w$-dimensional state vector $\mathbf{X_i}$ denotes the dynamics of each unit $i$ where $i=1,2,....N$. The dynamics of each isolated node are assumed to be identical and described by  $F: \mathbb{R}^w \rightarrow \mathbb{R}^w $. $f_p:\mathbb{R}^{2\times w}\rightarrow \mathbb{R}^w$ and $f_h:\mathbb{R}^{3\times w}\rightarrow \mathbb{R}^w$ denote the conjugate coupling functions for the $1$-simplex and $2$-simplex interactions respectively. 
 $\epsilon_p$ illustrates the coupling strength for the pairwise interactions and $\epsilon_h$ for the triadic ones. $C$ is the $N \times N$ adjacency matrix  associated to the $1$-simplex, where $C_{ij} = 1$, if $i^{th}$ and $j^{th}$ node are connected and $C_{ij} = 0$, otherwise. Similarly, $D$ is the $N \times N \times N$ adjacency tensor associated to the $2$-simplex where $D_{ijk} = 1$ if the nodes $i^{th}$, $j^{th}$ and $k^{th}$  are connected and $D_{ijk} = 0$, otherwise.
 Here the adjacency tensor can be defined  as $D_{ijk}= C_{ij}C_{jk}C_{ki}$. 

We consider a globally-connected network of $N$ Stuart-Landau ($SL$) limit cycle oscillators, interacting with each other through $1$-simplex and $2$-simplex interactions. The equations governing the dynamics of the system are:
	\begin{eqnarray}
		\dot{x_i}&=&(1 - x_i^2 - y_i^2)x_i - wy_i  + \frac{\epsilon_p}{N} \sum_{j=1}^{N} C_{ij}(y_j - x_i) \nonumber\\
		&&+\frac{\epsilon_h}{N^2} \sum_{j=1}^{N}\sum_{k=1}^{N} D_{ijk}(x_j y_k^2 - x_i^3) \nonumber\\
		\dot{y_i}&=&(1 - x_i^2 - y_i^2)y_i  + wx_i  + \frac{\epsilon_p}{N}  \sum_{j=1}^{N}C_{ij}(x_j - y_i)\nonumber\\
		&& + \frac{\epsilon_h}{N^2} \sum_{j=1}^{N}\sum_{k=1}^{N} D_{ijk} (y_j x_k^2 - y_i^3) 
		\label{eq:eqmodel}  
	\end{eqnarray}
	with $i=1,2,...N$. $\omega$ is the intrinsic frequency of the  oscillators.  The $N$ nodes in the network are coupled to each other by `conjugate' variables. 
 
 

 The dynamical equations (Eq.~\ref{eq:eqmodel}) have been solved numerically using the $RK4$ method with a step-size of $0.01$ after removing transients of the order of $10^5$. The initial conditions were randomly chosen within the interval $ [-1,1]$. \par To characterize the nature of the transition  from oscillatory state to death state in our model system, we adopt an order parameter in terms of average amplitude \cite{resmi2012amplitude}
 \begin{equation}
  A(\epsilon)=\frac{a(\epsilon)}{a(0)}   \nonumber
 \end{equation}
where $a(\epsilon) = \sum_{i=1}^{N}\frac{1}{N} (\langle x_{i,max}\rangle_t -\langle x_{i,min} \rangle_t)$ with $\langle ... \rangle_t$ representing the average over long time. $a(\epsilon)$  is a measure of the difference of the global maxima and minima of the time series data of all the oscillators, computed over a long time after discarding a sufficient amount of transients at a particular coupling strength $\epsilon$. $a(0)$ is the average amplitude at $\epsilon=0$.  The normalized order parameter $A(\epsilon)$ is greater than $0$  when the coupled system  is in the oscillatory state. However, when the system attains a steady state, i.e., $A(\epsilon) = 0$. This order parameter has been calculated adiabatic by varying the value of $\epsilon$ ($\epsilon_p$ or $\epsilon_h$) in forward and backward continuation.  This process is termed `adiabatic' since the system's state in the previous run is used as the initial conditions of the next run of the simulations \cite{verma2017explosive}. During the forward continuation process, the coupling strength $\epsilon$ increases slowly in steps of $\delta\epsilon$. When a specified value of $\epsilon$ is reached, the backward continuation process starts where $\epsilon$ is now decreased slowly in a similar fashion and continued until the minimum value of $\epsilon$ is reached.  
\begin{figure}
\includegraphics[width=0.5\textwidth]{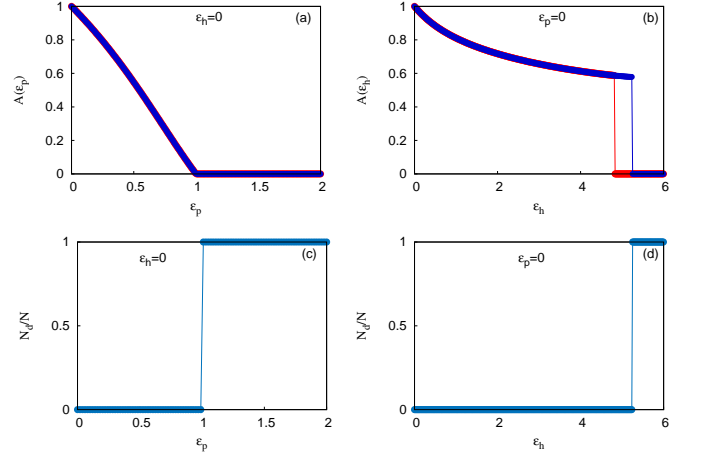}
\caption{\label{fig1}  (Color online) Amplitude order parameter $A$ 
(a) for $1$-simplex interactions  and (b) for $2$-simplex interactions. The fraction of oscillators $N_d/N$ is calculated for (c) $1$-simplex interactions and (d) $2$-simplex interactions. Here $\omega=1$ and $N=100$.} 
\end{figure}
\paragraph{Numerical Results for 1-simplex (dyadic) interactions:}
To investigate the dynamics with only $1$-simplex or pairwise interactions, we set $\epsilon_h$ = 0 in Eq.~(\ref{eq:eqmodel}). The coupling strength $\epsilon_p$ varies adiabatically in steps of $0.02$ in both forward and backward continuation. Fig.~\ref{fig1}(a) exhibits a second-order transition from oscillation to death of oscillations. The order-parameter $A(\epsilon_p)$ gradually decreases to zero, indicating suppression of oscillations in the system with a gradual increase in coupling strength. The forward and backward continuations show no change in the transition point, i.e., in both cases, the critical value of $\epsilon_p$ remains the same, as observed in Fig.~\ref{fig1}(a). Hence, in the presence of only conjugate dyadic links, the system shows a continuous and reversible second-order phase transition from oscillation to the death state, without any hysteresis via supercritical Hopf bifurcation. 

\begin{figure}
\includegraphics[width=0.45\textwidth]{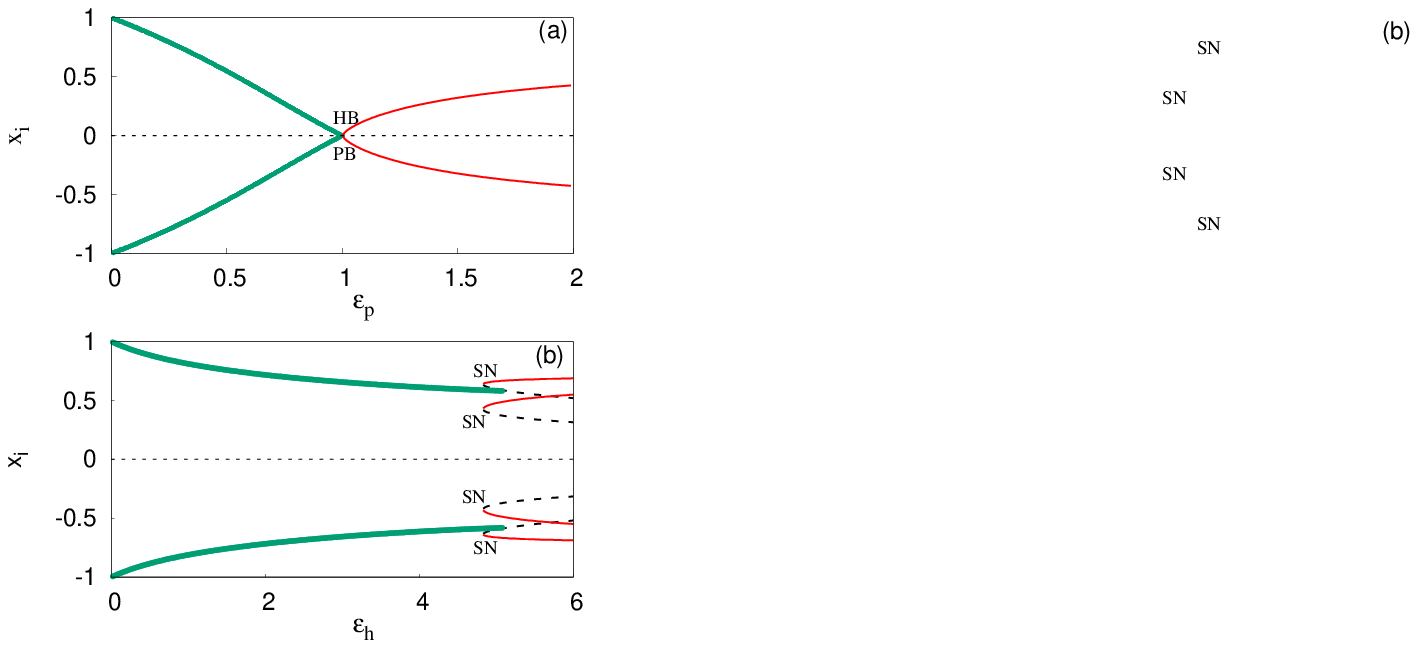}
\caption{\label{fig2} (Color online) The bifurcation diagram of the coupled $SL$ oscillators is plotted. (a) for $1$-simplex interactions and (b) for $2$-simplex interactions. The red ($\rho$) solid lines correspond to the stable fixed points, the dashed black lines correspond to the unstable ones, and the green ($\gamma$) solid lines indicate the oscillatory solutions from the stable limit cycle. $i=1,2.....N$.} 
\end{figure}

Next, we discuss the dynamics of the system at the transition points. The bifurcation diagram for the $SL$ oscillators with $1$-simplex interactions has been illustrated in Fig.~\ref{fig2}(a). Here, a supercritical Hopf bifurcation destroys the oscillations from the stable limit cycle, displayed in green ($\gamma$). A subsequent supercritical pitch-fork bifurcation also occurs immediately, giving rise to the two stable solutions which have been displayed in red ($\rho$) in Fig.~\ref{fig2}(a)).

\paragraph{Numerical results for 2-simplex (triadic) interactions:}
Next, we consider the case of $\epsilon_p =0$ in Eq.~\ref{eq:eqmodel}, i.e., only the triadic interactions are prevalent in the model. $\epsilon_h$ is varied in an adiabatic manner, from $0$ to $\epsilon_{h,max}$ in steps of $\delta \epsilon_h =0.02$ for the forward continuation. During this process, the magnitude of the order parameter $A(\epsilon_h)$ suddenly drops to $0$, indicating an explosive transition from oscillatory to steady state, as shown in Fig.~\ref{fig1}(b). The transition point from oscillatory to death state occurs for a critical $\epsilon_h = 5.2$. For the backward continuation process, we decrease $\epsilon_h$ from $\epsilon_{h,max}$ to $0$ similarly and observe that the amplitude order-parameter jumps from $0$ to non-zero values at around $\epsilon_h=4.82$. Hence, it is clear that the forward and the backward transition points do not coincide. This indicates a hysteresis region in the parameter space, i.e., a bi-stable region where both oscillatory and HSS states co-exist. Such an abrupt and sudden transition with a hysteresis region can be categorized as a first-order transition.

To investigate whether all the oscillators are going to death state simultaneously or some oscillators may attain death state faster than others, we define a parameter $N_d/N$ where $N_d$ is the number of oscillators going to steady state, and $N$ is the total number of oscillators. This parameter is also computed in a similar adiabatic way of forward continuation, as described previously. Fig.~\ref{fig1}(c) and Fig.~\ref{fig1}(d) depict the abrupt jump from $0$ to $1$ for $1$-simplex and $2$-simplex interactions respectively. This indicates that the oscillators attain a death state simultaneously after crossing a threshold value of $\epsilon_h$. We note that this value of $\epsilon_h$ agrees well with our results for forward continuation. 

The dynamics at the transition point for the $2$-simplex interactions differ from that of the $1$-simplex ones. The oscillations of the system remain in complete synchronization till a critical value of $\epsilon_h$ is reached. At this point, the system stabilizes to newly created fixed points through saddle-node bifurcations. Due to the bifurcations, two pairs of stable and unstable fixed points appear. The oscillations from the stable limit cycle continue and collide with the unstable fixed points. This leads to the disappearance of the limit cycle, and the system settles down to any of the stable fixed points, depending on the initial conditions. This scenario of a stable limit cycle being destroyed by an unstable fixed point is the mechanism of a sub-critical Hopf bifurcation. The entire dynamics of the model system at the transition point are illustrated in Fig.~\ref{fig2}(b). 

\begin{figure}
\includegraphics[width=0.45\textwidth]{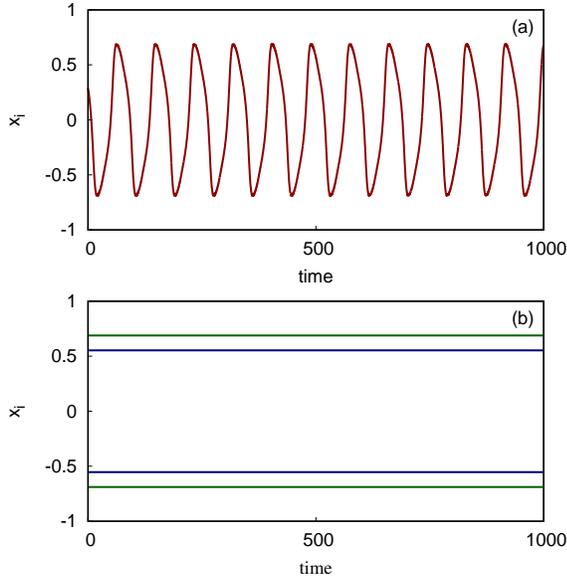}
\caption{\label{fig3} (Color online) Time series of the coupled $SL$ oscillators with $2$-simplex interactions is plotted. (a) at $\epsilon_h=2.45$, i.e., in the oscillatory region (b)  at $\epsilon_h=6.11$, i.e., at steady state region,  one set of fixed points ($x_i\approx\pm 0.55)$ which are displayed in dark purple ($\Psi$) and the other set of fixed points ($x_i\approx\pm 0.68)$ which are displayed in dark green ($\Gamma$) also appear depending on the choice of initial conditions for the same value of $\epsilon_h=6.11$. The other parameters are $\omega=1$, $N=100$ and $i=1,2.....N$.} 
\end{figure}

\begin{figure}
\includegraphics[width=0.45\textwidth]{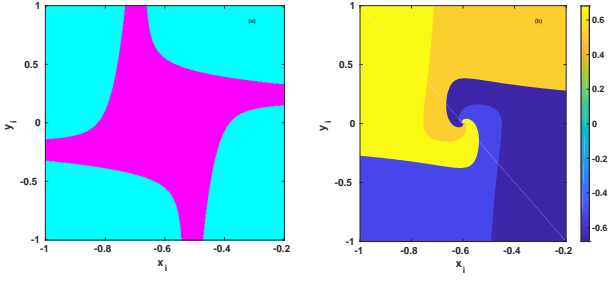}
\caption{\label{fig4}(Color online) Basin of attraction illustrating the stable states in the system. (a) At the hysteresis region $\epsilon_h=5.0$. The pink $(\phi)$ color represents the initial conditions for the oscillatory state and light blue ($\Lambda$) for the homogeneous steady state (b) at the death region $\epsilon_h=5.6$. The four distinct colors represent the four different choices of fixed points for the system at a steady state. Here, $N=100$, $\omega=1$. $i=1,2.....N$} 
\end{figure}

Fig.~\ref{fig3} demonstrates  the transition of the model system from oscillation to oscillation-quenching. The system shows periodic and completely synchronous oscillations (Fig.~\ref{fig3}(a)) for coupling strengths below the critical value of $\epsilon_h$ for which the transition to a steady state takes place. The system goes to a fixed point as $\epsilon_h$ increases beyond this point. Depending on the choice of initial conditions, the oscillators may settle down to any of the four homogeneous steady states shown in Fig.~\ref{fig3}(b). Fig.~\ref{fig3}(b) exhibits the fixed points $x_i^* \approx \pm 0.55$, displayed in dark purple ($\Psi$) and  $x_i^* \approx \pm 0.68$, illustrated in dark green ($\Gamma$). It is interesting to note that the system has a kind of mirror-image symmetry. That is to say, the dynamics occurring in the phase space above the origin are mirrored in the phase space below the origin. Fig.~\ref{fig4}(a) displays the basin of attraction at the hysteresis region. The numerical results have been calculated for $\epsilon_h=5.0$ since the hysteresis region stretches from $\epsilon_h\approx 4.82$ to $\epsilon_h\approx 5.25$. Here, we confirm the presence of bi-stability through the two regions illustrated in the figure. The pink ($\phi$) region corresponds to the initial conditions triggering the oscillatory state, and the light blue ($\Lambda$) region denotes the ones for the death state. The initial conditions $x_i$ and $y_i$ have been varied from $-1$ to $1$. In Fig.~\ref{fig4}(b), the basin of attraction for the four fixed point attractors in the system has been displayed. The value of the coupling strength has been fixed at around $5.6$ since we are exploring the basin after the system achieves a steady state. In this figure, each of the four differently shaded (colored) regions is populated by the initial conditions for the four different fixed points that the oscillations settle down to during the steady state. 
In Fig.~\ref{fig4}, we consider $x=x_i$ and $y=y_i$ since the system is in either complete synchronization or HSS in the hysteresis region. 

\begin{figure}
\includegraphics[width=0.45\textwidth]{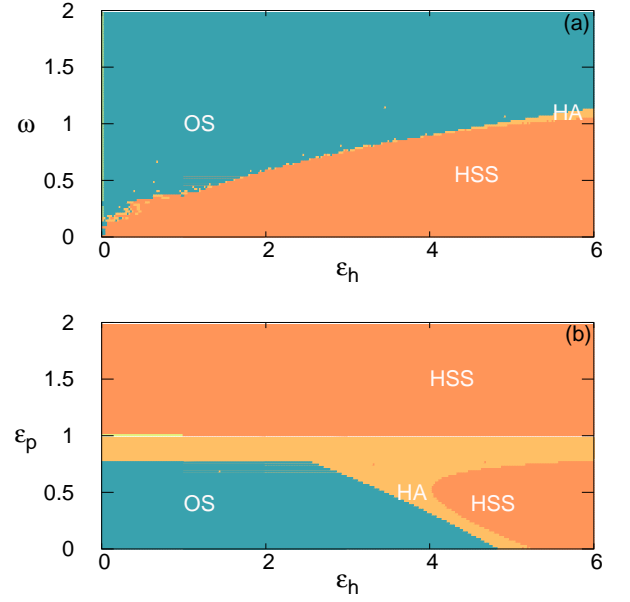}
\caption{\label{fig5} (Color online) Different dynamical domains of $N$ coupled $SL$ oscillators are plotted in the parameter plane (a) $(\epsilon_h-\omega)$ for $\epsilon_p = 0$ and (b) $(\epsilon_h-\epsilon_p)$ for $\omega=1$. Here OS, HA, and HSS denote oscillatory state, hysteresis area, and homogeneous steady state.}
\label{phase}
\end{figure}

We further explore the impact of the parameter $\omega$ on the dynamics of the $SL$ oscillators with $2$-simplex interactions. The parameter-space ($\epsilon_h-\omega$) diagram Fig.~\ref{fig5}(a) illustrates that the transition from oscillatory to steady state occurs for $\omega \le 1$ via first-order transition. It is important to note that here in HSS, the system may stabilize to any of the four fixed points, as depicted in Fig.~\ref{fig3}(b).

\paragraph{Both 1-simplex and 2-simplex interactions:}
We also study a case where $SL$ oscillators interact through $1$-and $2$-simplex interactions. 
For a weak $1$-simplex interaction ($\epsilon_p \le 0.8$), the system manifests a first-order transition from the oscillatory to the steady state with an increase in  
$\epsilon_h$. For intermittent values,  $1<\epsilon_p<0.8$, a region of bi-stability or hysteresis is observed regardless of $\epsilon_h$ values. The system with stronger $\epsilon_p \ge 1.0$ always attains a steady state and has no additional effect of $2$-simplex interaction.
Hence, we can conclude that the 2-simplex conjugate interactions lead to more complex dynamics with the first-order transition from oscillatory to steady state in the globally coupled Stuart-Landau oscillators as compared to second-order transition in the presence of $1$-simplex conjugate interaction.  

\paragraph{Analytical Calculations:}
Analysing Eq.~\ref{eq:eqmodel} and putting $\epsilon_h=0$ along with $\epsilon_p=\epsilon$ we derive the following stable fixed point: \begin{eqnarray}
		x^*  &= \sqrt\frac{{1 - \epsilon + \omega -\frac{\omega}{\epsilon}+ \eta}}{2} \nonumber \\
		y^* &= \frac{\epsilon(\epsilon -\omega)(\epsilon + \omega)\sqrt{1+\omega+ \frac{\epsilon^2 \eta -\epsilon(\epsilon^2 + \omega)}{\epsilon^2}}}{\sqrt{2} \eta \epsilon^2}
	\end{eqnarray}
	where $\eta= \frac{\sqrt{\epsilon^2 (\epsilon -\omega)^3 (\epsilon +\omega)}}{\epsilon^2}$. To find the condition of the transition from oscillatory to death state,  the Jacobian $\mathscr{W}$ is written in the form of block circulant matrices \cite{davis1979circulant}, i.e., $\bf{\mathcal{J}}=circ(\zeta_1,\zeta_2,...\zeta_2)$. \par $\zeta_1 =$ 	$\begin{bmatrix}
		1 - \epsilon - 3x^2 -y^2 & \frac{\epsilon}{2} - \omega-2xy
  \\
		\frac{\epsilon}{2} + \omega-2xy &	1 - \epsilon -x^2 -3y^2  \\	
	\end{bmatrix}$ and $\zeta_2$ = $\begin{bmatrix}
	0 & \frac{\epsilon}{2} \\
	\frac{\epsilon}{2} & 0
\end{bmatrix}$
 A $2(N-1)$ degeneracy exists for $2(N-1)$ eigenvalues of $\mathscr{W}$. These eigenvalues are equal to the eigenvalues of the matrix $(\zeta_1 - \zeta_2)$. The other eigenvalues of $\mathscr{W}$ will equal to those of the matrix $\zeta_1+ (N-1) \zeta_2$.
 The characteristic equation corresponding to the matrix $\zeta_1 + (N-1)\zeta_2$ is given by \begin{eqnarray}
\lambda^2 + d_1 \lambda +d_0 =0
\end{eqnarray}
where $d_1 = -2 + 2\epsilon + 4x^{*^2}+4y^{*^2} $ and $d_0= 1-2\epsilon+\omega^2-4x^{*^2}+4\epsilon x^{*^2}+3x^{*^4}+4\epsilon x^{*}y^{*} -4y^{*^2}+6x^{*^2}y^{*^2}+3y^{*^4}$. The condition for the Hopf bifurcation point is derived  according to Routh-Hurwitz criterion \cite{liu1994criterion} by putting $d_1=0$. Hence, the critical value of $\epsilon$ for the transition of the system from the oscillatory to OD state through Hopf bifurcation is given by, \begin{eqnarray}
	\epsilon_{HB} = \frac{1}{3}\Big(-1+2\sqrt{1+3\omega^2}\Big) \label{eq:hopf}
\end{eqnarray}

The condition for pitchfork bifurcation may also be derived by putting $d_0=0$. \begin{eqnarray}
	\epsilon_{PB} = \frac{1}{2}(1+\omega^2) \label{eq:pb}
\end{eqnarray}
Putting $\omega=1$ in Eq.~\ref{eq:hopf} and Eq.~\ref{eq:pb}, $\epsilon_{HB}$ and $\epsilon_{PB}$ are both found to be $1$, which is in exact agreement with our numerical results. 

We analyze Eq.~\ref{eq:eqmodel} again, and this time, by considering only the $2$-simplex interactions, i.e., $\epsilon_p=0$ and $\epsilon_h$ is generalized to $\epsilon$.  Numerical simulations reflect that the coupled system stabilizes at the HSS. Hence the backward transition point for the explosive transition can be computed by the stability analysis of HSS. The stable fixed points for this system are, $ x_i^* = x^* , y_i^* = y^* , \forall i = 1,...,N $ , where, 
\begin{eqnarray}
x^* &= \pm\frac{1}{2}\sqrt{\frac{{2+\epsilon + \alpha \pm \sqrt{2}(1+\epsilon)\beta}}{1+\epsilon}} \\ \nonumber
y^* &= \pm\frac{1}{4(1+\epsilon)w}( \pm \epsilon \pm \alpha +\sqrt{2} \epsilon \beta +\sqrt {2}\epsilon^2 \beta)x*
\end{eqnarray}
where $\alpha = \sqrt{\epsilon^2 - 4(1 + \epsilon)w^2}$ and $\beta = \sqrt\frac{{\epsilon + 2w^2 + 2\epsilon  w^2 - \sqrt{\epsilon^2 - 4(1 + \epsilon)w^2}}}{\epsilon(1 + \epsilon)^2}$. $2N \times 2N$ Jacobian matrices can be written in the form of block circulant matrices $\bf{\mathscr{J}=circ(\mathscr{M}_1,\mathscr{M}_2,...\mathscr{M}_2)}$.

\par $\mathscr{M}_1 = $ 
$\begin{bmatrix}
			1 -\Omega x^2 + \frac{y^2}{3}\kappa & -\omega + \frac{2}{3} \kappa x y  \\
			\omega + \frac{2}{3} \kappa x y &	1 -\Omega y^2 + \frac{x^2}{3} \kappa \\	
		\end{bmatrix} $ 
and $\mathscr{M}_2$ = $\begin{bmatrix}
\frac{\epsilon y^2}{3} & \frac{2 \epsilon xy}{3} \\
\frac{2\epsilon xy}{3} & \frac{\epsilon x^2}{3}
\end{bmatrix}$ with $\Omega=  3(1+\epsilon)$ and $\kappa = (-3+\epsilon)$

As explained earlier, there will be $2(N-1)$ eigenvalues of $\mathscr{J}$ with $2(N-1)$ degeneracy. These eigenvalues are equal to the eigenvalues of the matrix $(\mathscr{M}_1 - \mathscr{M}_2)$. The rest of the eigenvalues will be equal to the eigenvalues of the matrix $\mathscr{M}_1+ (N-1) \mathscr{M}_2$ and can  be analyzed to know the stability of the HSS state. The characteristic equation corresponding to this matrix is given by \begin{eqnarray} 
\lambda^2 + c_1 \lambda + c_0 =0
\label{charac_eqn}
\end{eqnarray} 
where $c_1 =  -2 + 4x^{*^2} + 2\epsilon x^{*^2} + 4 y^{*^2} + 2\epsilon y^{*^2}$ and $c_0= 1 +\omega^2 -4 x^{*^2} - 2 \epsilon x^{*^2} + 3x^*{^4} - 3\epsilon^2 x^{*^4} - 4y^{*^2} - 2\epsilon y^{*^2} + 6 x^{*^2} y^{*^2} + 24 \epsilon x^{*^2} y^{*^2} + 6 \epsilon^2 x^{*^2} y^{*^2} + 3y^{*^4} - 3\epsilon^2 y^{*^4}$. Solving this Eq.~\ref{charac_eqn}, we have two eigenvalues:
\begin{eqnarray}
	\lambda_{1,2}= \frac{1}{2(1+\epsilon)}\Big(-2-2\epsilon-\epsilon^2-2\alpha-\epsilon\alpha\pm\gamma\Big)
\end{eqnarray}
where $\gamma =  \sqrt{ 2(2-4\alpha+8\omega^2 +\epsilon(4+16\omega^2-6\alpha)+\epsilon^2 \xi)}$,  where $\xi = 2+ 6\omega^2 +\epsilon ( \alpha +\epsilon -2\omega^2)$ and $\alpha= \sqrt{\epsilon^2 -4(1+\epsilon)\omega^2}$
Equating the real parts of the eigenvalues to zero, we obtain the expression for the backward transition point through saddle-node bifurcation,  \begin{eqnarray}
	\epsilon_{SN}=2(\omega^2+\sqrt{\omega^2+\omega^4}) \label{eq:sn}
\end{eqnarray}
    Hence, we confirm that this model system stabilizes to a homogeneous steady state through a saddle-node bifurcation. The calculated backward transition point from Eq.~\ref{eq:sn} after considering $\omega=1$ is found to be 4.82, which agrees with our numerical results. 

\paragraph{Conclusion}
Here, we have investigated the $SL$ oscillators coupled through conjugate variables and having higher-order interactions. In a number of real-world systems modeling, coupling through a dissimilar variable is desired.  Examples of such situations include certain electronic circuits \cite{singla2011exploring} and epidemiological models \cite{goleniewski1996modelling}. Conjugate coupling is also found in laser experiments, where a photodiode detector monitors the light emitted by each laser diode through feedback \cite{kumar2008stable} and also in ecological settings, where  predators consume prey from external systems and push the food web to new steady states \cite{karnatak2014conjugate}. 
In ecological systems, it is difficult to map the co-existence in communities with multiple species without resorting to the concept of higher-order interactions. Cross-predation is a common phenomenon in ecosystems. Hence, one potential application of conjugate coupling with higher-order interactions is to study the stabilization of the food web in ecosystems characterized by frequent cross-predation among multiple species.

We report the first-order transition, an irreversible  transition from oscillation to steady state by considering $2$-simplex conjugate-coupled interactions instead of the reversible and continuous transition for $1$-simplex conjugate interactions only. As expected, triadic connections yield more complex dynamics  with the coexistence of oscillatory and steady states in the transition region. 
We have observed a thin hysteresis region in the parameter space. This region contains two stable states; oscillatory and steady. This bi-stability in the hysteresis region has important applications in the natural systems where oscillatory and steady states have been found to coexist, such as fruit flies in eclosion rhythm \cite{beuter2003nonlinear} and injection of stimulus pulse to restart the activity in the sino-arterial node \cite{guevara1992three} among others. After the transition to the steady state, the system stabilizes at four different fixed points, depending on the choice of the initial conditions. We analyzed the basins of attraction in the hysteresis and steady-state regions and confirmed our numerical results analytically by performing linear stability analysis. The analytical calculations have further allowed us to derive the backward transition point that matches the numerical results.  We have reported the occurrence of consecutive saddle-node bifurcations at the transition region, leading the system to settle down to HSS. 

It should be interesting to extend the present study 
for other network topologies, e.g., Erd\"{o}s-R\'{e}yni random networks and small-world networks, etc. Further, this study is restricted to only dyadic and triadic interactions. The inclusion of higher-order simplicial structures  is a straightforward extension of  the present work. To conclude, the study may be used to understand the origin of  hysteresis and oscillation suppression phenomena in biological, ecological, and chemical systems, which have multiple agents interacting with each other and need to be mapped with higher-order interactions. 

\paragraph{Acknowledgements}
RG acknowledges financial support from UGC New Delhi for SJSGC fellowship. SJ gratefully acknowledges SERB Power grant SPF/2021/000136. MDS acknowledges financial support from SERB, Department of Science and Technology (DST), India (Grant No. CRG/2021/003301).
\bibliography{main.bib}
\end{document}